\documentclass{llncs}

\usepackage{cite}
\usepackage[english]{babel}
\usepackage[utf8]{inputenc}
\usepackage{amsmath}
\usepackage{graphicx}
\graphicspath{{imgs/}{../imgs/}}
\usepackage{epstopdf}
\usepackage{subcaption}

\usepackage[hidelinks]{hyperref}
\hypersetup{
  colorlinks = true,
  urlcolor   = blue,
  linkcolor  = blue,
  citecolor  = red
}

\AtBeginDocument{\renewcommand\tablename{Tab.}}

\title{Simulation of an Optional Strategy in the Prisoner's Dilemma in Spatial and Non-spatial Environments \thanks{The final publication is available at Springer via \url{http://dx.doi.org/10.1007/978-3-319-43488-9_14}}}

\author{Marcos Cardinot \and Maud Gibbons \and Colm O'Riordan \and Josephine Griffith}

\institute{Dept. of Information Technology, National University of Ireland, Galway, Ireland \\
\email{marcos.cardinot@nuigalway.ie}}

\index{Cardinot, Marcos}
\index{Gibbons, Maud}
\index{O'Riordan, Colm}
\index{Griffith, Josephine}

\begin{document}
\maketitle

\begin{abstract}
This paper presents research comparing the effects of different environments on
the outcome of an extended Prisoner's Dilemma, in which agents have the option
to abstain from playing the game. We consider three different pure strategies:
cooperation, defection and abstinence. We adopt an evolutionary game theoretic
approach and consider two different environments: the first which imposes no
spatial constraints and the second in which agents are placed on a lattice
grid. We analyse the performance of the three strategies as we vary the loner's
payoff in both structured and unstructured environments. Furthermore we also
present the results of simulations which identify scenarios in which
cooperative clusters of agents emerge and persist in both environments.
\end{abstract}

\section{Introduction}

Within the areas of artificial life and agent-based simulations, evolutionary
games such as the classical Prisoner's Dilemma \cite{Axelrod1984,Smith1982},
and its extensions in the iterated form, have garnered much attention and have
provided many useful insights with respect to adaptive behaviours. The
Prisoner’s Dilemma game has attained this attention due to its succinct
representation of the conflict between individually rational choices and
choices that are for the better good. However, in many social scenarios that we
may wish to model, agents are often afforded a third option --- that of
abstaining from the interaction. Incorporating this concept of abstinence
extends the Prisoner’s Dilemma to a three-strategy game where agents can not
only cooperate or defect but can also choose to abstain from a game
interaction. There have been a number of recent studies exploring this type of
game \cite{Xia2015,Ghang2015,Jeong2014,Hauert2008,Izquierdo2010}.

In addition to analysing the evolution of different strategies and different
outcomes, previous work has also explored the effect of imposing spatial
constraints on agent interactions. Traditionally, these studies assume no such
constraints and agents are free to interact with all other agents in well-mixed
populations \cite{Axelrod1984}. However, many models consider restricting
interactions to neighbourhoods of agents on some pre-defined topology. These
more expressive models include lattices \cite{Nowak1992,Hauert2003}, cycles
and complete graphs \cite{Jeong2014}, scale-free graphs \cite{Xia2015} and
graphs exhibiting certain properties, such as clustering coefficients
\cite{Menglin2013}.

In this paper we adopt an evolutionary approach to evolve populations of agents
participating in the extended Prisoner’s Dilemma \cite{Nowak2006}.
We consider two different environmental settings: one with no enforced
structure where agents may interact with all other agents; and another in which
agents are placed on a lattice grid with spatial constraints enforced, where
agents can play with their immediate eight neighbours (Moore neighbourhood). In
both environmental settings, an agent's fitness is calculated as the sum of the
payoffs obtained through the extended Prisoner's Dilemma game interactions.  We
investigate the evolution of different strategies (cooperate, defect and
abstain) in both spatial and non-spatial environments. We are particularly
interested in the effect of different starting conditions (number of different
strategies and placement of different strategies) and the different values for
the loner’s payoff ($L$) on the emergence of cooperation. We identify
situations where the simulations converge to an equilibrium, where no further
changes occur. These equilibria can be fully stable (no change) or quasi-stable
(with a small cycle length).

The paper outline is as follows: In Section~\ref{sec:related} an overview of work
in the extended game and of spatial evolutionary game theory is presented.
Section~\ref{sec:method} gives an overview of the methodology employed. In
Section~\ref{sec:nonspatial}, we discuss the non-spatial environment. We firstly
present an analysis of pairwise interactions between the three pure strategies.
Secondly, evolutionary experiments using all three strategies are presented.
Thirdly, we explore the robustness of a population of cooperative and
abstaining strategies when a defecting strategy is added to the population. In
Section~\ref{sec:spatial}, we discuss the environment where agents are placed on
a lattice grid, in which their interactions are constrained by their local
neighbourhood. Again an analysis of pairwise interactions is first undertaken
followed by an exploration of the outcomes when all three strategies are
randomly placed on the grid. Based on these findings, we explore different
starting groupings of the three strategies, i.e. placed in a non-random manner
on the grid. This will allow identification of starting configurations that
lead to stable cooperation.

\section{Related Work}
\label{sec:related}

Abstinence has been studied in the context of the Prisoner's Dilemma (PD) since
Batali and Kitcher, in their seminal work \cite{Batali1995}, first introduced
the optional variant of the game. They proposed the opt-out or ``loner's''
strategy, in which agents could choose to abstain from playing the game, as a
third option, in order to avoid cooperating with known defectors. Using a
combination of mathematical analysis and simulations, they found that
populations who played the optional games could find routes from states of low
cooperation to high states of cooperation. Subsequently, as this extension has
grown in popularity and renown, optional participation has been successfully
incorporated into models alongside other cooperation enhancing mechanisms such
as punishment \cite{Hauert2008} and reputation \cite{Olejarz2015,Ghang2015},
and has been applied to probabilistic models \cite{Xia2015}.

The study of optional participation can be broadly separated into approaches:
one that directly incorporates abstinence into the traditional PD game (the
loner's strategy), and another known as conditional cooperation. Models that
incorporate the loner's strategy treat the option to abstain as an alternative
strategy  for agents to employ \cite{Batali1995,Jeong2014}, separate to the
option to cooperate or defect. These models tend to be more grounded in
mathematical models with less of an emphasis on experimental simulations, which
often-times have been shown to produce unexpected results \cite{Hauert2003}. On
the other hand, conditional cooperation models
\cite{Arend2005,Joyce2006,Izquierdo2010}, also known as conditional
disassociation, incorporate abstinence into cooperation strategies. These
models lend themselves more easily to Axelrod-style tournaments
\cite{Axelrod1984}. They tend to focus on exit options or partner-leaving
mechanisms, and often lack a spatial aspect, which has since been shown to
increase the number of abstainer strategies thus increasing the chances of
cooperation evolving \cite{Jeong2014}.

The work that most closely resembles our own is that of Hauert and Szab\'o
\cite{Hauert2003}. They consider a spatially extended PD and public goods game
(PGG), where a population of $N$ agents are arranged and interact on a variety
of different geometries, including a regular lattice. Three pure strategies
(cooperate, defect and abstain) are investigated using an evolutionary
approach. Results showed that the spatial organisation of strategies affected
the evolution of cooperation, and in addition, they found that the existence of
abstainers was advantageous to cooperators, because they were protected against
exploitation. However, there exists some major differences between their model
and the one proposed here. Hauert and Szab\'o focus on a simplified PGG as
their primary model for group interactions, and separately use the PD only for
pairwise interactions. In our model, agents interact by playing a single round
of the PD with each of their neighbours. Additionally, Hauert and Szab\'o
focused on one set of initial conditions for their simulations, using a fixed
ratio of strategies. Our work explores a wider range of initialization settings
from which we gleam more significant insights, and identify favourable
configurations for the emergence of cooperation.

\section{Methodology}
\label{sec:method}

In order to explore these strategies and, in particular, the effect of
introducing abstinence, we propose a set of experiments in which each agent
randomly plays a number of one-shot, two-person extended Prisoner's Dilemma
game. An evolutionary approach is adopted with a fixed-size population where
each agent in the population is initially assigned a fixed strategy. Fitness is
calculated and assigned based on the payoffs obtained by the agents from
playing the game. Simulations are run until the population converges on a
single strategy, or configuration of strategies.

In the traditional Prisoner’s Dilemma game there are four payoffs corresponding
to the pairwise interaction between two agents. The payoffs are: reward for
mutual cooperation ($R$), punishment for mutual defection ($P$), sucker’s
payoff ($S$) and temptation to defect ($T$). The dilemma arises due to the
following ordering of payoff values: $S < P < R < T$. When extending the game
to include abstinence, a fifth payoff is introduced, the loner’s payoff ($L$)
is awarded to both participants if one or both abstain from the interaction.

The value of $L$ should be set such that: (1) it is not greater than $R$,
otherwise the advantage of not playing will be sufficiently large to ensure
that players will always abstain and (2) it is greater than $S$, otherwise
there are no benefits to abstaining. This enables us to investigate the values
of $L$ in the range $[S,R]$, which in turn contrasts with the definition used
by Hauert and Szab\'o \cite{Hauert2003} who define abstainers as strategies who
perform better than groups of defectors but worse than groups of mutually
cooperating strategies. In their model, abstainers receive a payoff less than
$R$ and greater than $P$. We choose to explore a more exhaustive range of
values. The payoffs for the extended Prisoner's Dilemma game are illustrated in
Tab.~\ref{payoffs} and are based on the standard values used by Axelrod
\cite{Axelrod1984}.

\begin{table}
    \caption{Prisoner's Dilemma Game Matrix.}
    \label{payoffs}
    \begin{subtable}{.5\linewidth}
        \centering
        \setlength{\tabcolsep}{10pt}
        \begin{tabular}{c| c c c}
                           & {\bf C} & {\bf D} & {\bf A} \\
            \hline {\bf C} & R,R     & S,T     & L       \\
                   {\bf D} & T,S     & P,P     & L       \\
                   {\bf A} & L       & L       & L       \\
        \end{tabular}
        \caption{Extended game matrix.}
    \end{subtable}
    \begin{subtable}{.5\linewidth}
        \centering
        \setlength{\tabcolsep}{8pt}
        \begin{tabular}{c|c}
            {\bf Payoff} & {\bf Value} \\
            \hline
            {\bf T}      & $5$         \\
            {\bf R}      & $3$         \\
            {\bf P}      & $1$         \\
            {\bf S}      & $0$         \\
            {\bf L}      & $]0,3[$     \\
        \end{tabular}
        \caption{Payoff values.}
    \end{subtable}
\end{table}

As we aim to study the behaviour of agents in different scenarios, our first
model allows all agents to potentially interact (Sect.~\ref{sec:nonspatial}).
Our second model places topological constraints on the agent population which
restricts the potential interactions that can take place
(Sect.~\ref{sec:spatial}). This allows for the comparison between spatial and
non-spatial environments and allows us to identify similarities and differences
in conditions that promote cooperation. For both environments, two common sets
of experiments are considered:

\begin{enumerate}
    \item Pairwise comparisons: The
        abstainer strategies compete with one of the other strategies;
        firstly, an equal number of cooperators ($C$) and abstainers ($A$); and
        secondly, an equal number of defectors ($D$) and abstainers ($A$).
    \item All three strategies present: We adopt an unbiased environment in which
        initially each agent is designated as a cooperator ($C$), defector ($D$)
        or abstainer ($A$) with equal probability.
\end{enumerate}

Moreover, to further explore the effect of adding the option of abstinence, a
third experiment is undertaken in the non-spatial environment, where we seed
the population with a majority of one type of strategy (abstainers) and explore
if the population is robust to invasion from (1) a cooperator and (2) a
cooperator and a defector (Sect.~\ref{sec:nonspatialRobustness}). In order to
explore the effect of different initial spatial configurations, we also
undertake a third set of experiments in the spatial environment, which provide
an insight in to the necessary spatial conditions that may lead to robust
cooperation (Sect.~\ref{sec:spatialRobustness}).

\section{Non-spatial Environment}
\label{sec:nonspatial}

In this section, we present results of the experiments in the non-spatial
environment and settings as described previously in Section~\ref{sec:method}.
We use a tournament selection with size 2.

\subsection{Pairwise Comparisons}

The simulations involving cooperators, $C$ and abstainers $A$, verified the
expected outcomes where the cooperators quickly spread throughout the
population resulting in complete cooperation. This can be shown to be correct
by calculating the difference in the payoffs each strategy receives:

\begin{equation*}
\begin{array}{lcl}
    P_C - P_A   &   =   &   (|C-1|R + |A|L) -(|C|L + |A-1|L) \\
                &   =   &   |C-1|(R-L)
\end{array}
\end{equation*}

As $R > L$, $P_C - P_A > 0 $ and thus  the cooperators always dominate. Our
simulations confirm this result.

When comparing $D$ and $A$ strategies and their payoffs, we see:
\begin{equation*}
\begin{array}{lcl}
    P_D - P_A   &   =   & |D-1|P + |A|L - |A+D-1|L \\
                &   =   & |D-1|P - |D-1|L          \\
                &   =   & |D-1|(P-L)
\end{array}
\end{equation*}

If $L = P$, then either defectors or abstainers may dominate at any stage. If
$L > P$, then abstainers dominate. If $L < P$ then defectors dominate.
Figure~\ref{DApanmicticGen50} illustrates this behaviour in simulations for
different values of $L$ with an initial equal population of defectors and
abstainers. For each simulation, 100 separate runs are undertaken and the
average of the numbers of each strategy present per run are averaged per
generation and plotted. It can be seen that when $L < P$, the defectors have a
selective advantage and dominate. At $L = P$, neither the defectors nor the
abstainers have a clear advantage. When $L > P$, the abstainers have the
selective advantage and they now dominate in the majority of cases. The above
calculations assume all players play all other players; our simulations
approximate this result.

\subsection{All Three Strategies}

\begin{figure}[tb]
    \centering
     \begin{subfigure}[t]{0.48\textwidth}
        \includegraphics[width=\textwidth]{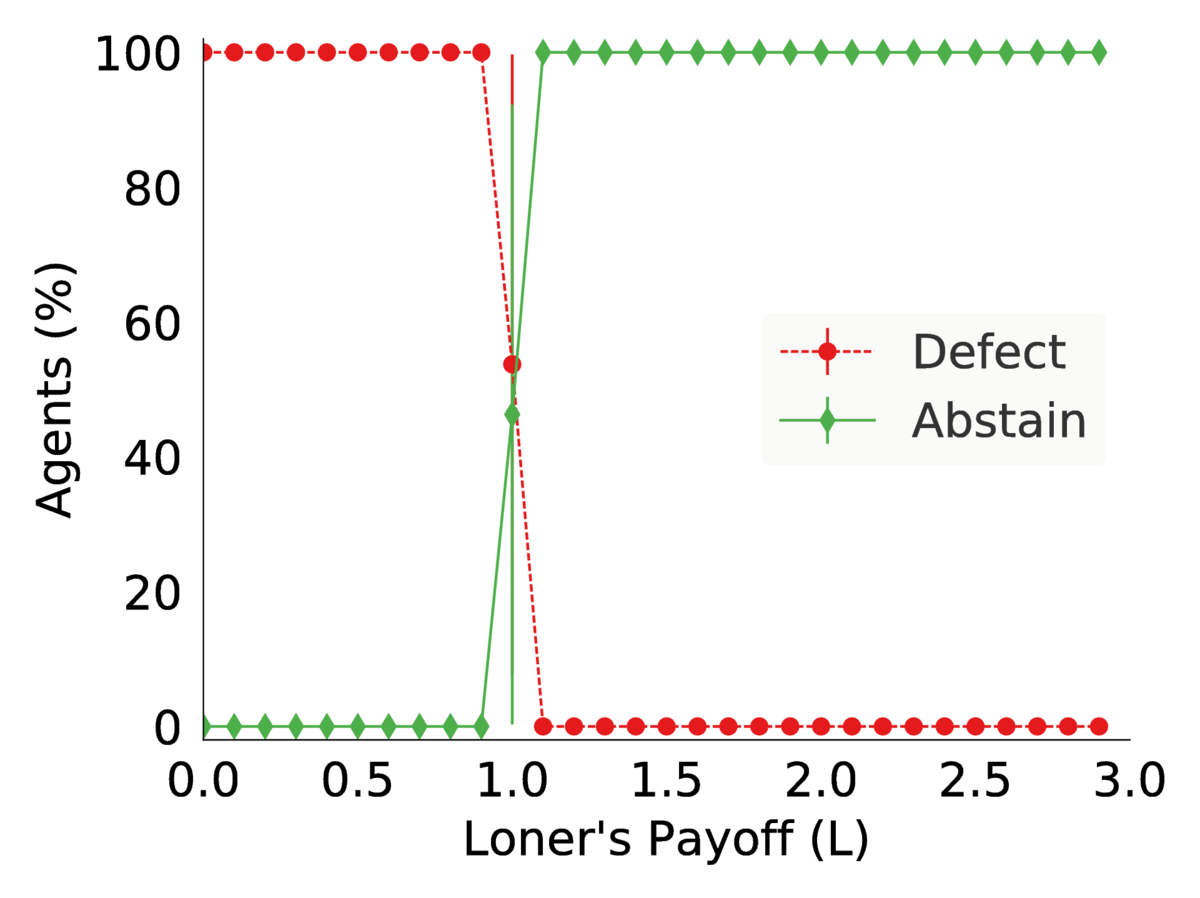}
        \caption{Generation 50 given an initial equal population of defectors
        and abstainers.}
        \label{DApanmicticGen50}
    \end{subfigure}
    ~
    \begin{subfigure}[t]{0.48\textwidth}
        \includegraphics[width=\textwidth]{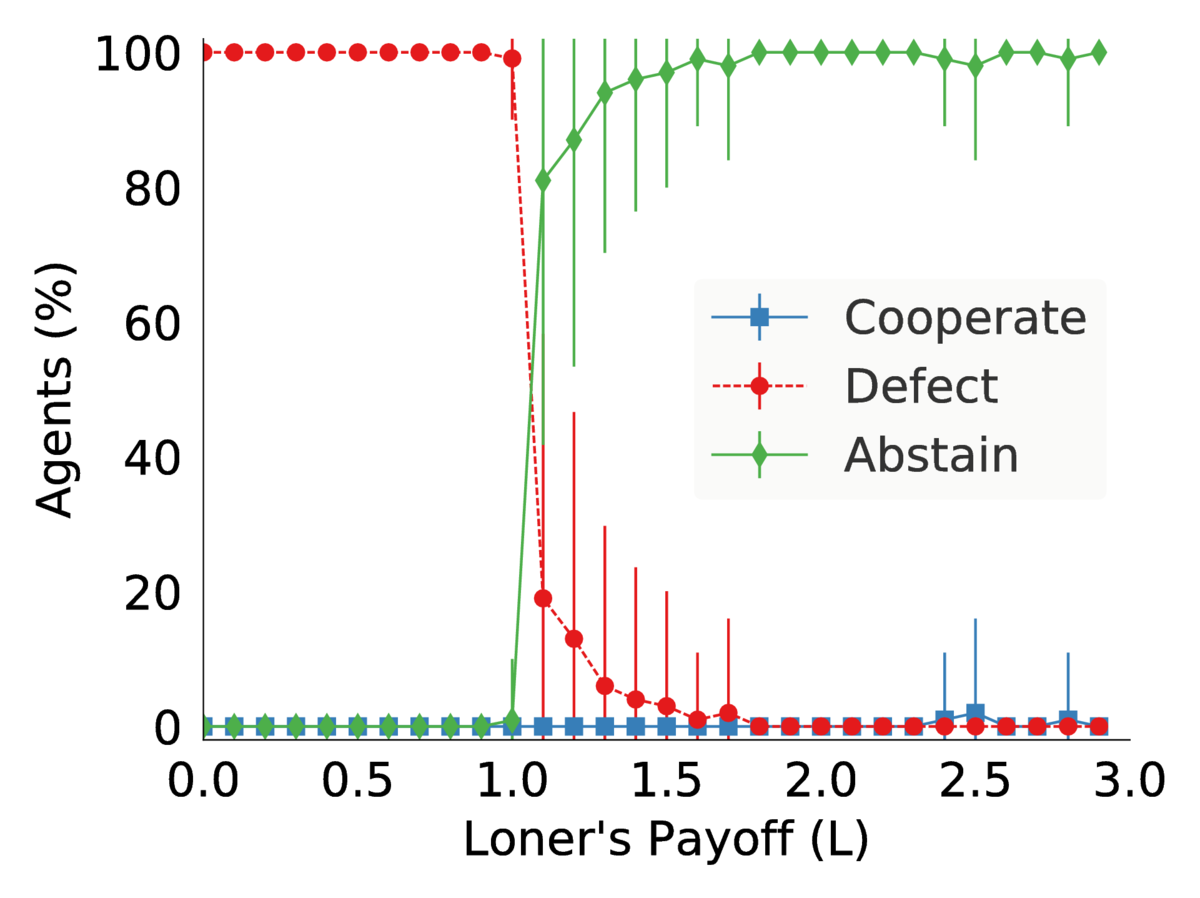}
        \caption{Generation 50 given an initial equal population of defectors,
            cooperators and abstainers.}
        \label{ACDpanmictic}
    \end{subfigure}
    ~
    \begin{subfigure}[t]{0.48\textwidth}
        \includegraphics[width=\textwidth]{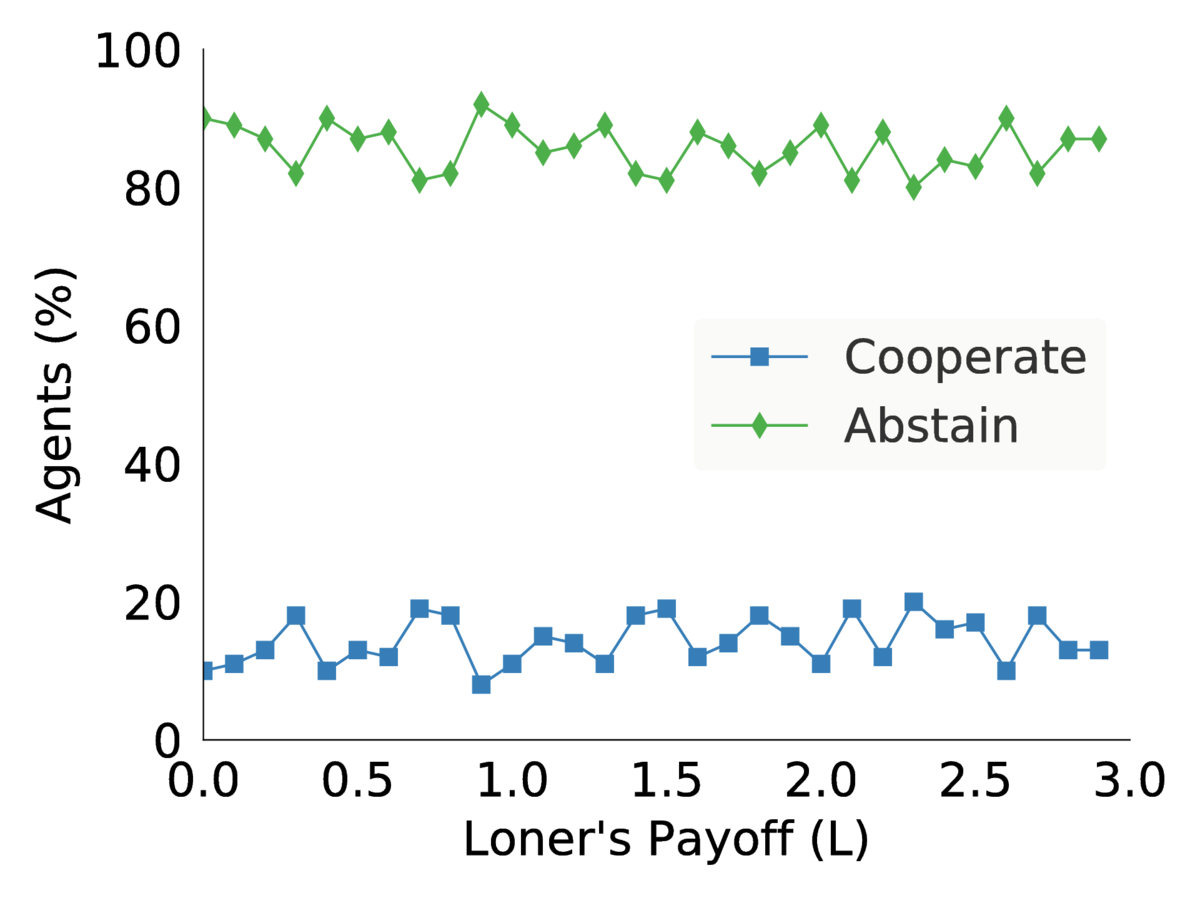}
        \caption{Robustness: initial non-spatial population of 1 cooperator
        and 99 abstainers.}
        \label{panmicticRobust1}
    \end{subfigure}
    ~
    \begin{subfigure}[t]{0.48\textwidth}
        \includegraphics[width=\textwidth]{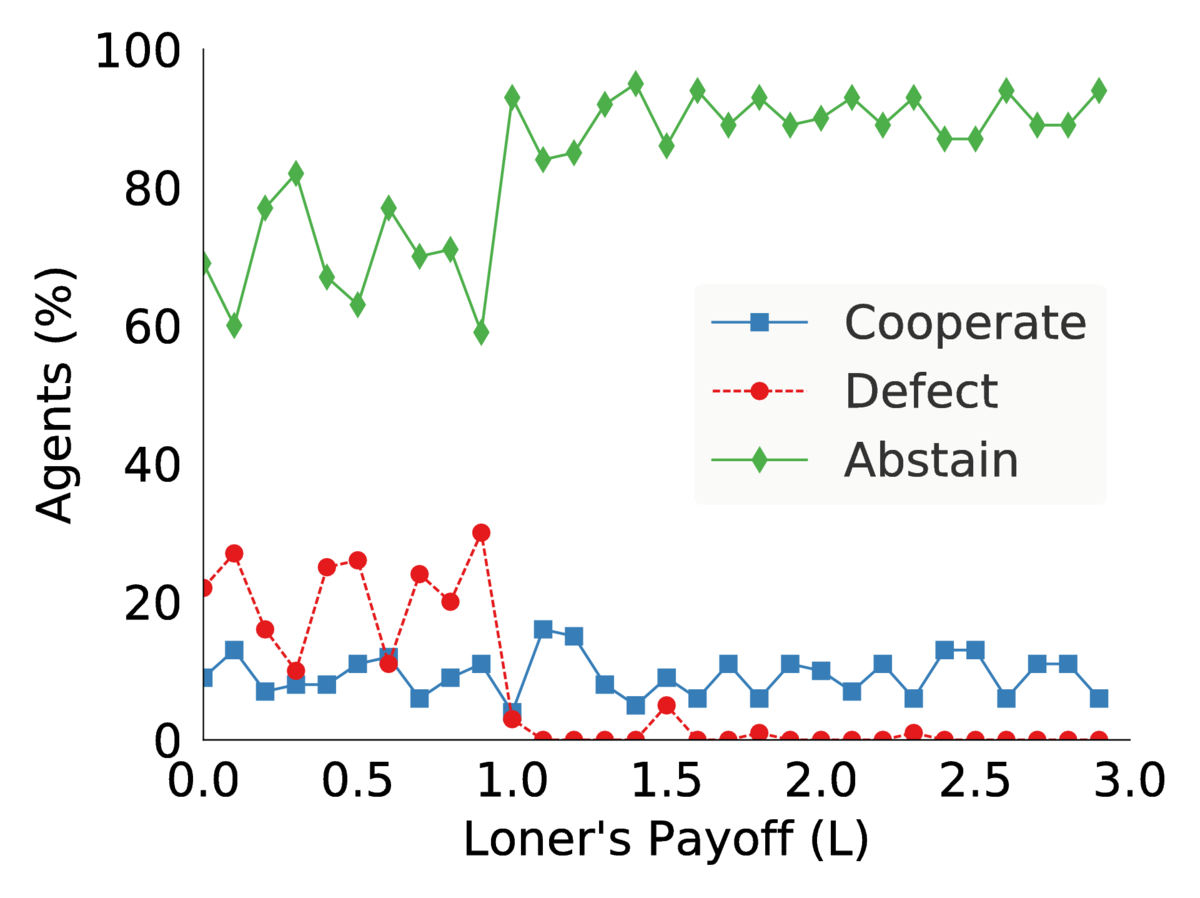}
        \caption{Robustness: initial non-spatial population of 1 cooperator,
        1 defector and 98 abstainers.}
        \label{panmicticRobust2}
    \end{subfigure}
    \caption{Experiments with a non-spatial population.}\label{fig:panmictic}
\end{figure}

In this experiment, an unbiased environment, with an initial population
consisting of the same number of cooperators (C), defectors (D) and abstainers
(A), is created. Figure~\ref{ACDpanmictic} illustrates the behaviour at
generation 50 across 100 individual runs. For $L < P$ defectors have already
dominated the population. For $L = P$, defectors still dominate but on a
minority of runs abstainers dominate. For $L > P$ this dominance of the
abstainers becomes more pronounced as the payoff for abstainers increases. In
fact, in some runs given the selective advantage of abstainers over defectors,
some cooperators outperform the defectors resulting in a fully cooperative run.

\subsection{Robustness}
\label{sec:nonspatialRobustness}

The previous experiments show the outcome for a range of starting conditions.
In this section, we explore the robustness of states to the introduction of a
defector. Initially, a population is created comprising one cooperator strategy
and the remainder of the strategies are all abstainers. In this situation, in
the first generation all strategies receive the same payoffs, $L$. Via
tournament selection, subsequent generations may comprise more than one
cooperative strategy. If this is the case, and these cooperative strategies are
chosen to play against each other, they receive a higher payoff than
abstainers, and cooperation will flourish. On the other hand, if the
cooperative strategy is not selected for subsequent generations, then the
population will consist only of abstainer strategies. This is illustrated in
Fig.~\ref{panmicticRobust1}, which shows the average of 100 runs. In any of
these runs, the evolutionary outcome is either a population comprising fully of
cooperators or a population comprising fully of abstainers. The value of $L$
does not affect this outcome as $L < R$ is always true.

In the second robustness experiment, the initial population consists of 98
abstainers, 1 cooperator and 1 defector. Figure~\ref{panmicticRobust2} shows
the outcomes after 50 generations. When $L < P$, as seen previously, the
defectors will have an advantage over abstainers. However, due to tournament
selection, there is a possibility that a defector will not be chosen for
subsequent generations. When $L > P$, the abstainers have the advantage over
the defectors given the possibility of mutual defection among defectors. The
defectors may continue to survive in the population given the presence of
cooperators whom they can exploit. We witness that the cooperators can still do
well given the benefits of mutual cooperation. However, the number of runs in
which cooperation flourishes is reduced due to the presence of defectors.  When
$L = P$, defectors and abstainers achieve the same payoff in their pairwise
interactions. However, defectors may do better in that they will exploit any
cooperators. As the cooperators die out, there is no selective advantage for
defectors but a level of robustness to invasion is observed.

In summary, these results show, when introducing one cooperator, abstainers and
cooperators can co-exist; but when adding one cooperator and one defector more
complex outcomes are possible.

\section{Spatial Environments}
\label{sec:spatial}

In this section, we are interested in exploring the larger range of outcomes
that result from the introduction of the spatial constraints. For the following
experiments, we replace the tournament selection used in the non-spatial
experiments with a mechanism whereby an agent adopts the strategy of the best
performing neighbour strategy. This is in line with standard approaches in
spatial simulations \cite{Nowak1992,Hauert2003}.

\subsection{Pairwise Comparison between Agents}

\begin{figure}[tb]
    \centering
    \includegraphics[width=.53\linewidth]{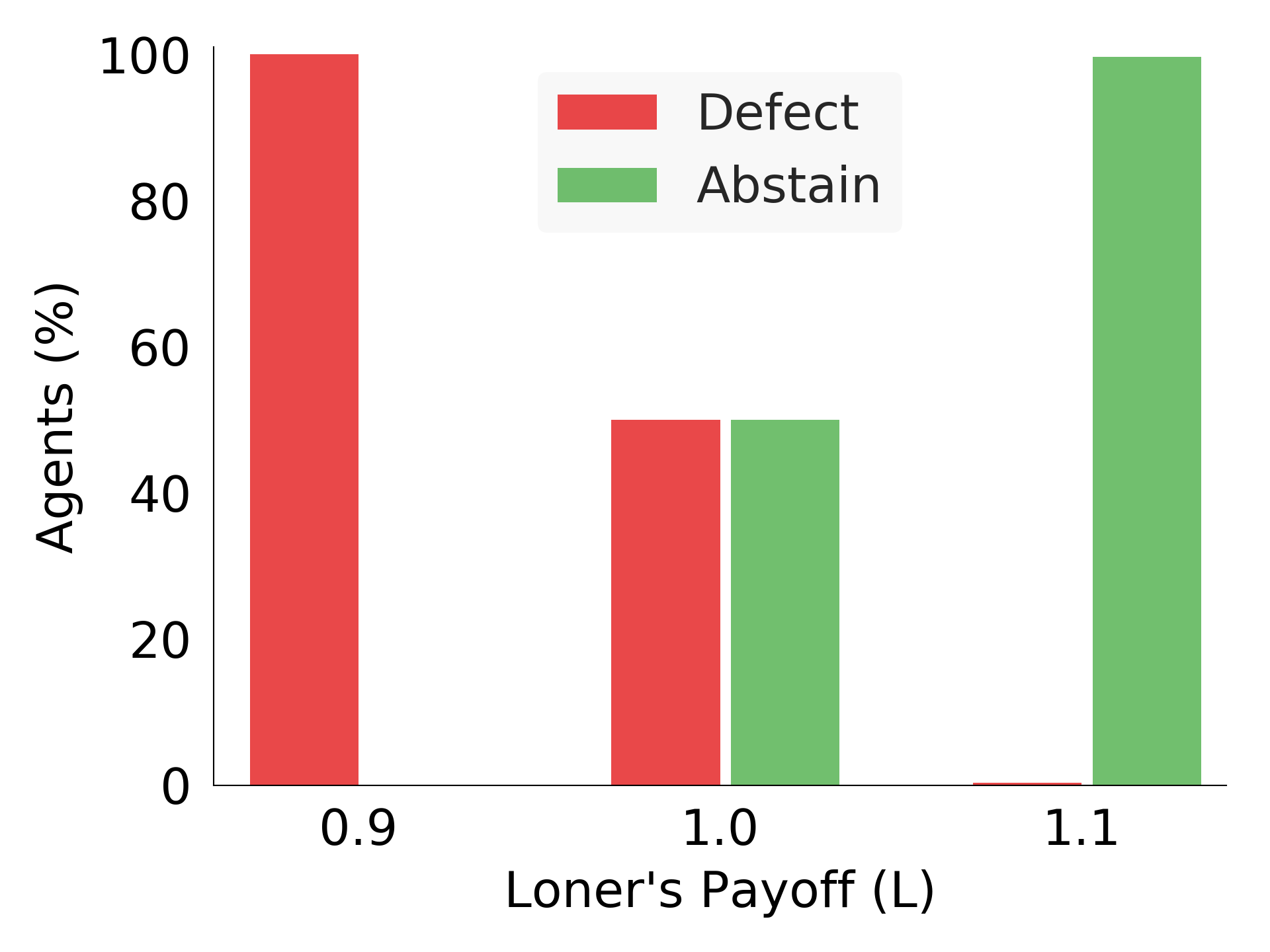}
    \caption{Percentage of defectors and abstainers at L = 0.9, 1.0 and 1.1.}
    \label{DApairwisegrid}
\end{figure}

When placing cooperator and defector agents randomly on the lattice grid, the
defecting agents will spread amongst the cooperators echoing previous findings.
When cooperator and abstainer agents are randomly placed on the grid, we find
that if there are at least two cooperators beside each other, cooperation will
spread, irrespective of the value of $L$ as cooperative agents playing with
each other will obtain a higher payoff than any adjacent abstainer agents.
Thus, neighbours will copy the cooperating strategy. Finally, when defector and
abstainer agents are randomly placed on the grid, we see from
Fig.~\ref{DApairwisegrid} that different outcomes occur depending on the value
of $L$. This is similar to the results observed in the non-spatial pairwise
comparison.

\subsection{All three strategies}
\label{sec:lattice_CDA}

In this experiment, equal numbers of the three strategies are placed
randomly on the grid. The outcome for $L < P$ is as expected with defectors
quickly dominating the population. However, in 65\% of simulations small
clusters of cooperators survive thanks to the presence of the abstainers. The
abstainers give the cooperators a foothold, allowing them to ward off invasion
from the defectors.

For $L = P$, defectors once again dominate, despite the tie, as they are able
to exploit cooperators in the population. Once again, some small groups of
cooperators survive with the same probability.

\begin{figure}[tb]
    \centering
     \begin{subfigure}[t]{0.31\textwidth}
        \includegraphics[width=\linewidth]{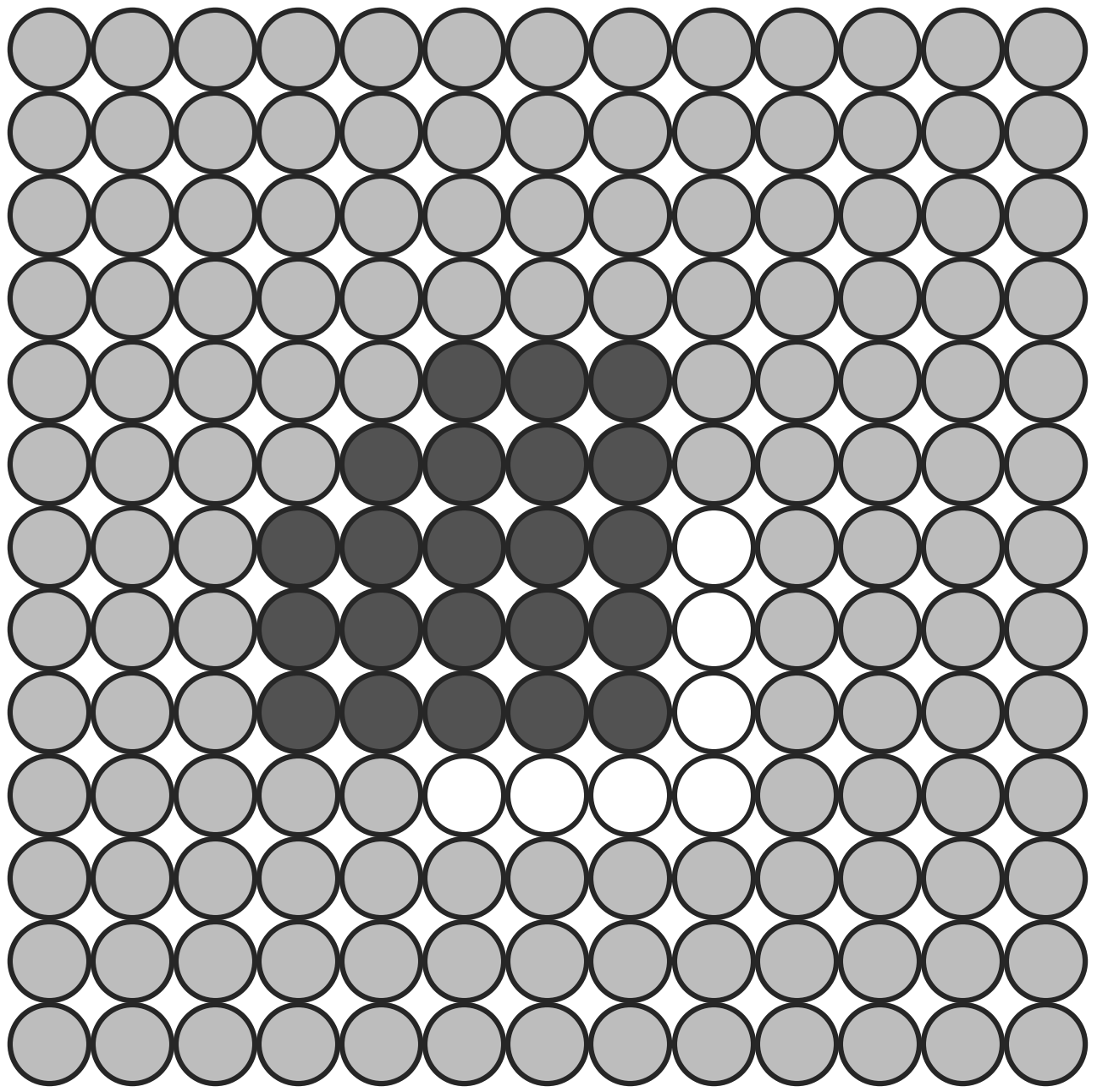}
        \caption{Glider (C and D)}
        \label{glider}
    \end{subfigure}
    ~
    \begin{subfigure}[t]{0.31\textwidth}
        \includegraphics[width=\linewidth]{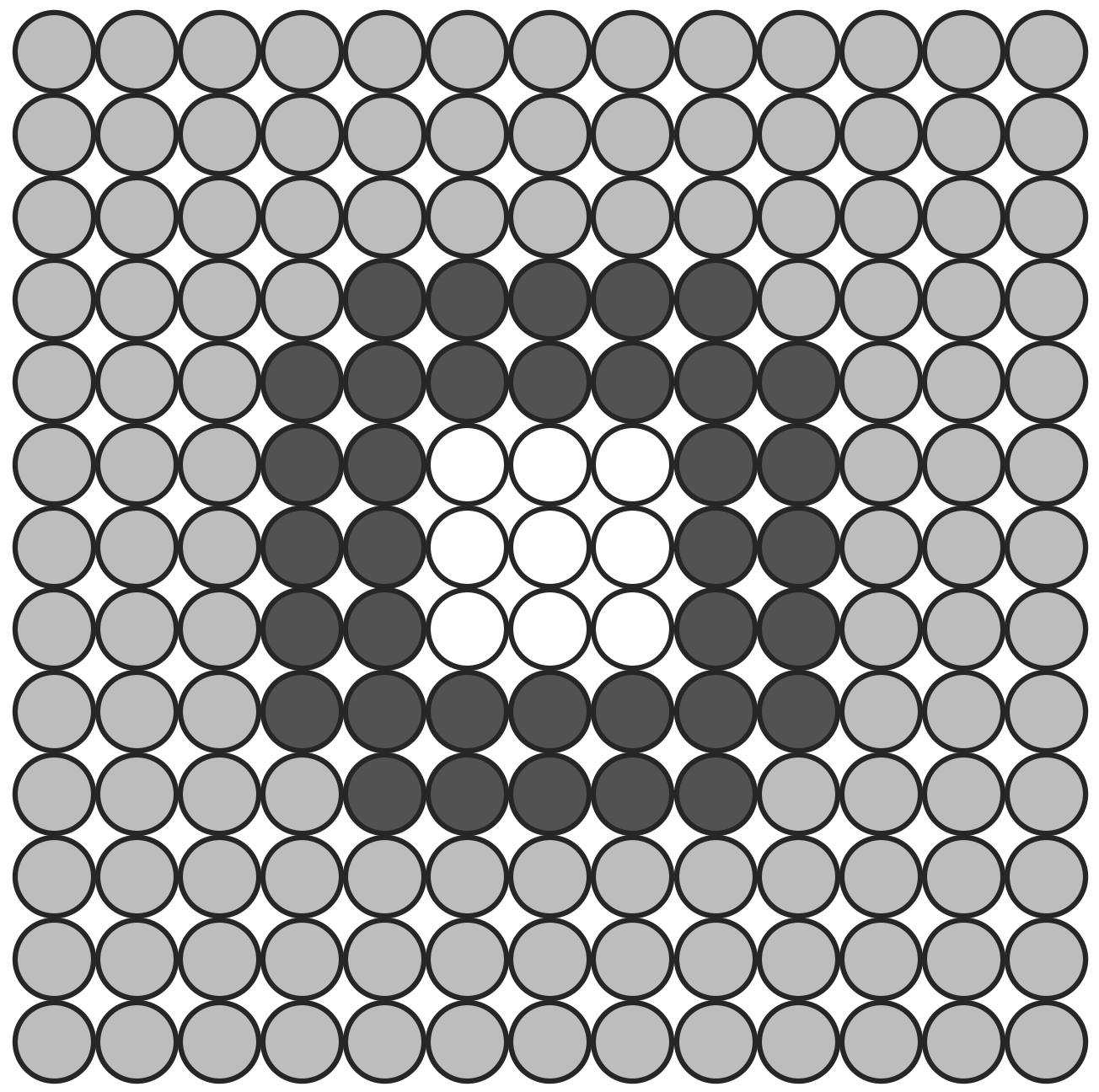}
        \caption{Cluster (C and D)}
        \label{cluster}
    \end{subfigure}
    ~
    \begin{subfigure}[t]{0.31\textwidth}
        \includegraphics[width=\textwidth]{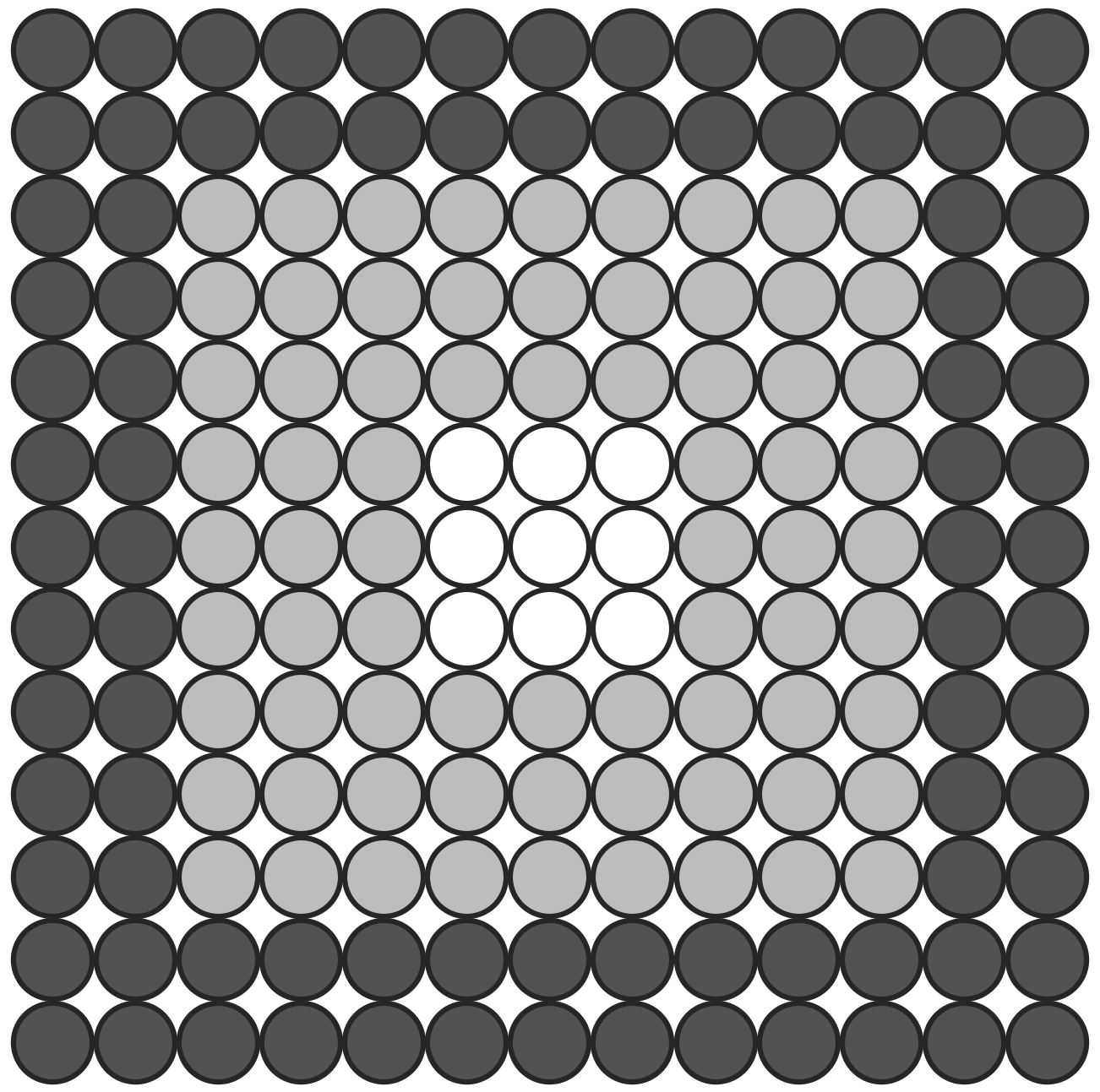}
        \caption{Initial configuration of C, A and D}
        \label{CAD}
    \end{subfigure}
    \caption{Experiments with a spatially organised population in a $100 \times 100$
    	lattice grid full populated with agents. The white dots represent cooperators (C),
        the black dots indicate defectors (D), and the grey dots are the abstainers (A).}
\end{figure}

A number of simulations are run varying $L$ from 1.1 to 2.0 where results show
similar emergent evolutionary stable patterns across all values of $L$ in this
range. There are two distinct outcomes; abstainers dominate; and abstainers
dominate with some sustained cooperation. Some level of cooperation is achieved
on average in 51.5\% of simulations for values of $L$ in the range $[1.1,2.0]$.
In these runs, a cooperative cluster (of minimum size 9), surrounded by
defectors, forms in the early generations and remains a stable feature in
subsequent generations. The presence of defectors, surrounding the cooperative
cluster, prevents the abstainers from being invaded by the cooperators.
Similarly the defector strategies remain robust to the spread of abstainers
given their ability to exploit the cooperators. In essence, a symbiotic
relationship is formed between cooperators and defectors.  Figure \ref{cluster}
shows a screenshot of a cooperator and defector cluster in a simulation where
abstainers have dominated. This configuration, once reached, is stable in these
settings.

As the value of $L$ increases we also witness newer phenomena. For $L = 1.5$
and $L = 2.0$, we see cycles between two states where some of the surrounding
defectors fluctuate from defector to abstainer and back again. We also see an
increase in the size and amount of clusters when they are formed. For $L=[1.7,
1.9]$, we see ``gliders'' \cite{Gardner1970} where a group of defectors flanked
by a row of cooperators seemingly move across the grid, as shown in
Fig.~\ref{glider}. In reality, the cooperators invade the abstainers, the
defectors invade the cooperators, and the abstainers in turn invade the
defectors.

\subsection{Exploration of the Effect of Different Initial Spatial Configurations}
\label{sec:spatialRobustness}

The aim of this experiment is to investigate different initial spatial settings
of cooperators, defectors and abstainers to further explain the results
witnessed in the previous experiment (Sect.~\ref{sec:lattice_CDA}). One
interesting outcome from the previous simulations involved a stable situation
where one strategy (inner) could survive in a cluster of the same strategies
due to being surrounded fully by another strategy (middle) which, in turn, is
itself surrounded fully by the third (outer) strategy (see Fig.~\ref{cluster}).
In this case, it appeared that the inner strategy needs the protection of the
middle strategy to avoid invasion by the outer strategy and that the middle
strategy in turn needs the inner strategy to avoid invasion by the outer
strategy. It was noted that for cooperators surrounded by defectors, a minimum
inner cluster size of 9 was needed in order for this outcome to emerge.

Given three strategies, we consider all six permutations with respect to the
placement of strategies in the three different positions of inner, middle and
outer with an inner cluster of size 9, a middle cluster comprising 3 layers
around the inner cluster, and the remaining outer portion of the grid
containing only the third strategy. We label these six spatial configurations
according to the first letter of the strategy (C, D, A) and their initial
position (inner, middle, outer). Figure~\ref{CAD} is an illustration of the
initial conditions for the ``C-A-D'' spatial configuration.  We note that given
any initial configuration the outcome will not vary. This means that there is
no reason, other than for verification, to run a configuration multiple times.
Two values of $L$ are explored: $L < P$ and $L > P$ for each configuration. For
$L = P$, simulations reveal no selective pressure for interactions between
defectors and abstainers. These results involve a level of stochasticity which
do not give any meaningful insights, and thus are not further discussed in this
paper.

In every permutation of A, C, and D when $L < P$ the defectors dominate. Both
defectors and cooperators invade the abstainers and then the defectors begin to
invade the cooperators. We again observe that many clusters of cooperators, of
different sizes but of minimum size 9, remain robust to this invasion, as a
result to the presence of the abstainers. The initial placement of the
strategies dictates how many cooperative clusters are likely to remain robust
to invasion by defection.  Table~\ref{results} provides an overview of the
results from each scenario when $L > P$. The existence of the abstainer
strategies, in addition to the initial placement of the strategies, ensures
that defection will not dominate in all of the scenarios. In fact, in one
scenario (CAD), it results in a fully cooperative population.

\begin{table}
    \centering
    \caption{Results of Seeded Initial Settings.}
    \setlength{\tabcolsep}{5pt}
    \begin{tabular}{c| c| p{6.8cm}}
        \bf{Shape} & \bf{Outcome} & \bf{Description}\\
        \hline
        \bf{DCA}  & Defection spreads   & Abstainers are invaded. Clusters of cooperators survive amongst dominant defectors. \\
        \hline
        \bf{DAC}  & Defection spreads   & Similar outcome as above.\\
        \hline
        \bf{CDA}  & Structurally stable & A symbiotic cluster of cooperators and defectors persist among the abstainers.\\
        \hline
        \bf{CAD}  & Cooperation spreads & Abstainers buffer cooperators against defectors, allowing them to dominate.\\
        \hline
        \bf{ACD}  & Abstainers invaded  & Cooperators invade the inner abstainers to create a cluster resistant to defector invasion.\\
        \hline
        \bf{ADC}  & Abstinence spreads  & Clusters of cooperators, surrounded by defectors survive within the abstainer majority (see Fig.~\ref{cluster}). \\
    \end{tabular}
    \label{results}
\end{table}

We have seen in comparison in the non-spatial experiments that all strategies
may influence each other's payoffs and we observe a smaller set of outcomes.
When all the strategies are placed together, either defectors or abstainers
dominate. In the spatial scenarios, there are outcomes with robust clusters of
cooperators. In the robustness experiments, in the non-spatial scenarios, a
largely cooperative population is easily invaded by defectors; this is not the
case in the spatial scenario where we have shown that cooperators can be robust
to invasion for specific initial settings. In the non-spatial scenarios, with
the existence of abstainers, the population is largely robust and results in a
mixed equilibrium.

\section{Conclusions and Future Work}

In this paper, two different environments in which populations of agents played
an extended version of the Prisoner's Dilemma were considered: non-spatial
where all $N$ agents were potential partners for each other, and a population
organised on a lattice grid where agents can only play with their 8 immediate
neighbours. For both scenarios, three sets of experiments were performed: a
pairwise comparison of two strategies; experiments involving all three
strategies and an exploration of the conditions leading to cooperative
outcomes.

In the non-spatial environment, for the pairwise comparison, with agents
initially having equal number of cooperators and abstainers, cooperation
spreads throughout the population. The outcome when agents initially have an
equal number of defectors and abstainers is dependent on the loner's payoff
($L$). When all three strategies are initially equally present in the
population the value of the loner's payoff is again crucial. When the value of
$L$ is less than or equal to the punishment for mutual defection, the dominant
strategy is defection; in other cases abstinence spreads as a strategy and this
in turn can lead to cooperation spreading. In the robustness experiments, we
consider populations comprising of agents with abstainer strategies and explore
the effects of perturbing the population by the addition of firstly, a
cooperative agent and secondly agents with strategies of cooperation and
defection. Results show that only in the second scenario does the value of $L$
influence the outcome.

In the spatial experiments, similar outcomes arise for the pairwise
comparisons. When considering equal numbers of agents with all three strategies
some similarities between the spatial and non-spatial results are noticed, but
the spatial organisation allows for the clustering of cooperative agents. For
all values of the loner's payoff, defection dominates in addition to the
presence of some clusters of cooperators where these clusters are protected by
abstainers . As the loner's payoff increases above 1.5, the size of these
clusters of cooperative agents increases. In the experiments considering
different initial spatial configurations interesting behaviour was noted for
the six different possible starting initialisations. In all cases, irrespective
of the position of the cooperative strategies initially, and the value of $L$,
cooperative clusters persisted.

In previous work in the spatial Prisoner's Dilemma, it has been shown that
cooperation can be robust to invasion if a sufficiently large cluster of
cooperators form. However, given a random initialisation, this rarely happens
and defectors can dominate in most scenarios. With the introduction of
abstainers, we see new phenomena and a larger range of scenarios where
cooperators can be robust to invasion by defectors and can dominate.

Future work will involve extending our abstract model to more realistic
scenarios. There are many documented scenarios of symbiosis between entities
(individuals, species, plants and companies). In our simulations, we model
symbiosis between three distinct entities. We are interested in identifying
scenarios where insights obtained in our spatial configurations may apply; for
example, the planting of specific plants (abstainers) to prevent the invasion
one plant species (defector) into another native species (cooperator).

Future work will also involve performing a more detailed investigation into
emergent evolutionary stable patterns witnessed at different values of $L$ and
the exploration of other topologies with the goal of identifying structures
that allow robust cooperation.

\bigskip
\subsubsection*{Acknowledgments. }
This work is funded by CNPq–Brazil and the Hardiman Scholarship, NUI Galway.
The final publication is available at Springer via \url{http://dx.doi.org/10.1007/978-3-319-43488-9_14}.

\bibliographystyle{splncs03}
\bibliography{refs}

\begin{thebibliography}{10}
\providecommand{\url}[1]{\texttt{#1}}
\providecommand{\urlprefix}{URL }

\bibitem{Arend2005}
Arend, R.J., Seale, D.A.: {Modeling alliance activity: An iterated prisoner's
  dilemma with exit option}. Strategic Management Journal  26(11),  1057--1074
  (2005)

\bibitem{Axelrod1984}
Axelrod, R.M.: The evolution of cooperation. Basic Books, New York (1984)

\bibitem{Batali1995}
Batali, J., Kitcher, P.: Evolution of altruism in optional and compulsory
  games. J. Theor. Biol.  175(2),  161--171 (1995)

\bibitem{Gardner1970}
Gardner, M.: {Mathematical Games: The Fantastic Combinations of John Conway's
  New Solitaire Game Life}. Scientific American  223(4),  120--123 (1970)

\bibitem{Ghang2015}
Ghang, W., Nowak, M.A.: {Indirect reciprocity with optional interactions}. J.
  Theor. Biol.  365,  1--11 (2015)

\bibitem{Hauert2003}
Hauert, C., Szab{\"o}, G.: Prisoner's dilemma and public goods games in
  different geometries: Compulsory versus voluntary interactions. Complexity
  8(4),  31--38 (2003)

\bibitem{Hauert2008}
Hauert, C., Traulsen, A., Brandt, H., Nowak, M.A.: {Public goods with
  punishment and abstaining in finite and infinite populations}. Biol. Theory
  3(2),  114--122 (2008)

\bibitem{Izquierdo2010}
Izquierdo, S.S., Izquierdo, L.R., Vega-Redondo, F.: {The option to leave:
  Conditional dissociation in the evolution of cooperation}. J. Theor. Biol.
  267(1),  76--84 (2010)

\bibitem{Jeong2014}
Jeong, H.C., Oh, S.Y., Allen, B., Nowak, M.A.: Optional games on cycles and
  complete graphs. J. Theor. Biol.  356,  98--112 (2014)

\bibitem{Joyce2006}
Joyce, D., Kennison, J., Densmore, O., Guerin, S., Barr, S., Charles, E.,
  Thompson, N.S.: My way or the highway: a more naturalistic model of altruism
  tested in an iterative prisoners' dilemma. J. Artif. Soc. Soc. Simulat.
  9(2), ~4 (2006)

\bibitem{Menglin2013}
Li, M., O'Riordan, C.: The effect of clustering coefficient and node degree on
  the robustness of cooperation. In: Evolutionary Computation (CEC), 2013 IEEE
  Congress on. pp. 2833--2839. IEEE (2013)

\bibitem{Nowak2006}
Nowak, M.A.: Evolutionary Dynamics: Exploring the Equations of Life. Harvard
  University Press, Cambridge (2006)

\bibitem{Nowak1992}
Nowak, M.A., May, R.M.: Evolutionary games and spatial chaos. Nature
  359(6398),  826--829 (1992)

\bibitem{Olejarz2015}
Olejarz, J., Ghang, W., Nowak, M.A.: {Indirect Reciprocity with Optional
  Interactions and Private Information}. Games  6(4),  438--457 (2015)

\bibitem{Smith1982}
Smith, J.M.: Evolution and the theory of games. Cambridge University Press,
  Cambridge (1982)

\bibitem{Xia2015}
Xia, C.Y., Meloni, S., Perc, M., Moreno, Y.: Dynamic instability of cooperation
  due to diverse activity patterns in evolutionary social dilemmas. EPL
  109(5),  58002 (2015)

\end{thebibliography}

\end{document}